\documentclass[proof]{WileyASNA-v1}
\usepackage{hyperref}

\articletype{Proceedings}%

\received{26 April 2016}
\revised{6 June 2016}
\accepted{6 June 2016}

\begin{document}

\title{Needle in a Poissonian haystack --- \\ An X-ray astronomer’s guide to QPE fishing}

\author[1,2]{Erwan Quintin*}

\author[2]{Norman Khan}

\author[2]{Natalie A. Webb}

\author[2]{Robbie Webbe}

\author[3]{Richard D. Saxton}

\author[4]{Giovanni Miniutti}

\author[4]{Margherita Giustini}

\authormark{Erwan QUINTIN \textsc{et al}}

\address[1]{\orgdiv{European Space Astronomy Centre (ESAC)}, \orgname{European Space Agency (ESA)}, \orgaddress{\state{Madrid}, \country{Spain}}}

\address[2]{\orgdiv{Institute for Research in Astrophysics and Planetology (IRAP)}, \orgname{CNRS}, \orgaddress{\state{Toulouse}, \country{France}}}


\address[3]{\orgdiv{Telespazio UK for the European Space Agency (ESA)}, \orgname{European Space Astronomy Centre (ESAC)}, \orgaddress{\state{Madrid}, \country{Spain}}}

\address[4]{\orgdiv{Centro de Astrobiología (CAB)}, \orgname{CSIC-INTA}, \orgaddress{\state{Madrid}, \country{Spain}}}


\corres{*\email{erwan.quintin@esa.int}}

\presentaddress{European Space Agency (ESA), European Space Astronomy Centre (ESAC), Camino Bajo del Castillo s/n, 28692 Villanueva de la Cañada, Madrid, Spain}

\abstract{After six years of studies following the discovery of GSN069, a link is starting to appear between the elusive Quasi-Periodic Eruptions (QPEs) and other types of nuclear transients, among which are Tidal Disruption Events (TDEs). As such, observing strategies are adapting, with a current trend focusing on late-time X-ray follow-ups of (optical) TDEs. 

While these campaigns are so far proving quite successful, the inherent confirmation bias they introduce in our sample could lead the community to hasty, and perhaps erroneous, conclusions. It is thus important to still pursue the search for nuclear transients in other, more agnostic directions. In this work, we focus on the observational aspects of our field, and lay out two different methods that can be deployed in order to reveal new QPE sources. 

These complementary methods enable the detection of long-term ($\sim$years) and short term ($\sim$minutes) transient events, that would have otherwise been missed by the standard detection pipelines. Both of these methods can be used either for data mining in the 25 years worth of XMM-Newton archive, or to trigger real-time follow-ups upon a more recent discovery.}




\maketitle


\section{Introduction}
X-ray quasi-periodic eruptions (QPEs) are repeated soft bursts of thermal X-rays arising from the nucleus of some galaxies \citep[e.g][]{miniutti_nine-hour_2019}. These events are rare, the current sample comprising 12 sources. Their peaks last from an hour up to about a day, and repeat from every few hours to every 20 days. They show a strong observational correlation with another seemingly unrelated type of extragalactic nuclear transient, tidal disruption events \citep[TDEs, e.g.][]{gezari_tidal_2021}. These correspond to the destruction of an inbound star by the tidal forces around a super-massive black hole (SMBH). These events are rare as well, and the observed correlation has an infinitesimal chance of being a coincidence. There are different models attempting to explain the behaviour of QPEs and their correlation with TDEs, with one current leading model being that of a pre-existing Extreme Mass Ratio Inspiral \citep[EMRI, e.g.][]{amaro-seoane_relativistic_2018} with a $\sim$stellar-mass object orbiting around a SMBH, which is subsequently the host of an independent TDE from another star. This disruption will lead to the debris from the second star circularising around the central SMBH and forming a compact accretion disk; the EMRI, having survived the initial event, will collide with the disk twice per orbit, leading to the eruptions \citep[e.g.][]{linial_emri_2023}. 

While this model manages to explain most observable features of QPEs, some questions remain open, especially regarding the exact emission mechanisms, and the long-term evolution of these systems. Two different directions can be taken to answer these questions: either simulations, which require to account for the complex physics at play, and for the various spatial and temporal scales -- or, alternatively, increasing the current sample can allow to test the models by revealing trends, observing new behaviours, and in general providing more datapoints to test the models. In these proceedings, we are interested in the general directions an X-ray astronomer can take in their attempts to increase the existing sample of QPEs. For this purpose, we begin by assessing the current trends from the already existing sample (Sec.~\ref{sec1}), and underline the limits of these current approaches. We then suggest two alternative methods: one focused on the detection of X-ray TDEs (Sec.~\ref{sec2}), and another focused on a new approach to source detection specifically tailored to QPEs (Sec.~\ref{sec3}).

\section{Current methods of QPE detection, and their limits}
\label{sec1}
Naturally, the first stage to design new ways of detecting new QPE candidates is to assess how the current sample was found, and the limits of these current methods. The first-ever QPE source was GSN069 \citep{miniutti_nine-hour_2019}, observed in late 2018 as a potential X-ray TDE, owing to its large historical amplitude and soft spectrum. After this initial serendipitous finding, the discovery methods of QPEs fell into three categories. The first type of discovery method was through archival searches. The discovery of GSN069 sparked intense interest, and archival searches allowed to unearth RX~J1301.9+2747 \citep{giustini_x-ray_2020}, which had already been noticed as a peculiarly variable source \citep{sun_rx_2013}, but the discovery of GSN069 shed light on its true nature. A dedicated search for eruption-like features in XMM-Newton archival lightcurves led to the discovery of XMMSL1 J024916.6-041244 \citep{chakraborty_possible_2021}. A broader search for short-term variability, agnostic about the shape of the lightcurve, led to the discovery of half of a QPE-like eruption in AT2019vcb \citep{quintin_tormunds_2023}, which was later confirmed as a QPE source by additional detections by eROSITA \citep{bykov_further_2024}. The second method that has been implemented to detect QPEs is through blind searches, using the sky surveying capabilities of eROSITA. By looking for the specific temporal features over consecutives eRO-days that is expected from QPEs, five additional sources were found \citep[eRO-QPE1 to eRO-QPE5,][]{arcodia_x-ray_2021,arcodia_srgerosita_2025,arcodia_more_2024}. The third and final method of detection has leveraged the strong observational correlation between QPEs and TDEs, by organising intense X-ray monitoring campaigns at late times in the decay of (mostly optical) TDEs -- this has led to the detections of QPEs in AT2019qiz, AT2022upj, and ZTF19acnskyy \citep{nicholl_quasi-periodic_2024, chakraborty_discovery_2025, hernandez-garcia_discovery_2025}.

While the current methods have been proven successful in building the current sample of QPEs, they present some weaknesses that are worth keeping in mind. Archival searches have likely yielded most of their low-hanging fruits, as several studies have used various methodologies and exhausted the existing catalogs \citep[e.g.][]{chakraborty_possible_2021, webbe_searching_2023, quintin_tormunds_2023}. New attempts would have to go deeper than standard catalog products. On the other hand, blind search studies were mostly made possible by the unique capabilities and observing strategy of eROSITA -- now that it is offline, only the slews of XMM-Newton and the Einstein Probe could allow comparable detections, although with a lesser sensitivity, and without the short-term monitoring allowed by the consecutive eRO-days. And finally, looking for QPEs by following up on optical TDEs at late times has proven successful, but comes with three major drawbacks, which are that 1) the X-ray behaviour of TDEs is relatively unclear, and in particular if a given optical TDE is not brightening at late times in X-rays and there are no QPEs, the follow-up will result only in upper limits, which might not be considered as the most efficient use of precious exposure time of X-rays observatory, especially seeing that 2) it relies on the access to capable X-ray monitoring telescopes to perform the follow-up campaign, which is not guaranteed depending on the immediate future of the NICER and Swift missions; and finally 3) this method inevitably introduces both a selection bias and a confirmation bias into our sample of QPEs -- we will only detect QPE at late times in TDE if we only look for them there, and will miss the potential other sources of QPEs that might exist. As such, if both standard catalog-based archival data mining and eROSITA-enabled blind searches are not possible, we should try to design methods to find QPEs that attempt to alleviate the three issues underlined here for the TDE-based method. 

\section{First alternative: Following up on X-ray TDEs}
\label{sec2}
A first method we suggest here would focus on mitigating the first issue, which is the possibility of wasting follow-ups on X-ray--faint TDEs. A simple solution would be to only trigger a follow-up campaign in the case of an already-detected X-ray counterpart to a TDE. While most of the current TDE sample was optically detected (e.g. by ZTF), it is possible to detect X-ray bright transients, and classify them as TDEs based on their location (matches a known passive galaxy), X-ray spectrum (expected soft thermal emission), and X-ray luminosity (typically $10^{40}$-$10^{42}$~erg~s$^{-1}$) -- see for instance \citet{sacchi_supersoft_2023}. While detecting the X-ray counterpart to a TDE does not guarantee the presence of QPEs, it presents two advantages that will help justify an extensive monitoring campaign. First, it excludes the presence of dense absorbing materials along the line of sight, which could otherwise hinder or prevent the detection of QPEs. Second, it ensures that, even in the absence of QPEs, the X-ray exposure time is not wasted, as this valuable data can be used to better understand X-ray TDEs.

This method assumes that there is a way to blindly detect X-ray transients, and to somehow constrain their flux, spectral shape and location. We note that this is different from eROSITA's capacity to blindly detect QPEs, which relied on consecutive scans a few hours apart from each other. Several missions are able to provide such X-ray alerts. Einstein Probe was designed for this purpose explicitly \citep{yuan_einstein_2015}. The Swift observatory has developed a tool to perform such transient detections \citep{evans_real-time_2023}. With this idea in mind, we have developed a similar automatic transient detection system for the XMM-Newton observatory, which is implemented in the pipeline \citep{quintin_stonks_2024}, and complements the already-existing transient detection system in the slews of the satellite \citep{saxton_first_2008}. Similar philosophies should be kept in mind for future observatories \citep[e.g. NewAthena,][]{cruise_newathena_2025}. 

This suggested method however does not solve the issues of availability of X-ray monitoring telescopes, or that of the biases introduced in the sample. As such, we suggest an alternative method.


\section{Second alternative: Dedicated faint bursts search}
\label{sec3}
\begin{figure}[h]
    \centering
    \includegraphics[width=\columnwidth]{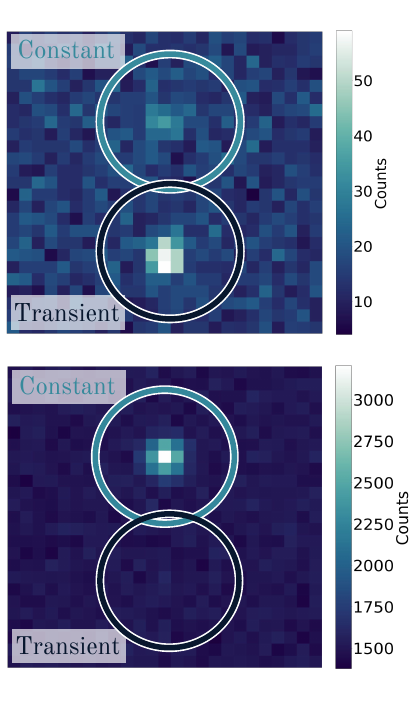}
    \caption{Illustration of the philosophy behind EXOD. The top panel shows a single frame, and the bottom panel an entire stacked observation. In both, we compare the appearance of a constant source (light blue circle) and a transient source (dark blue circle). This shows clearly that stacking over the whole exposure increases the signal-to-noise ratio of constant sources, but drowns out transient objects.}
    \label{fig:EXOD1}
\end{figure}

\begin{figure*}[h]
    \centering
    \includegraphics[width=\linewidth, page=1]{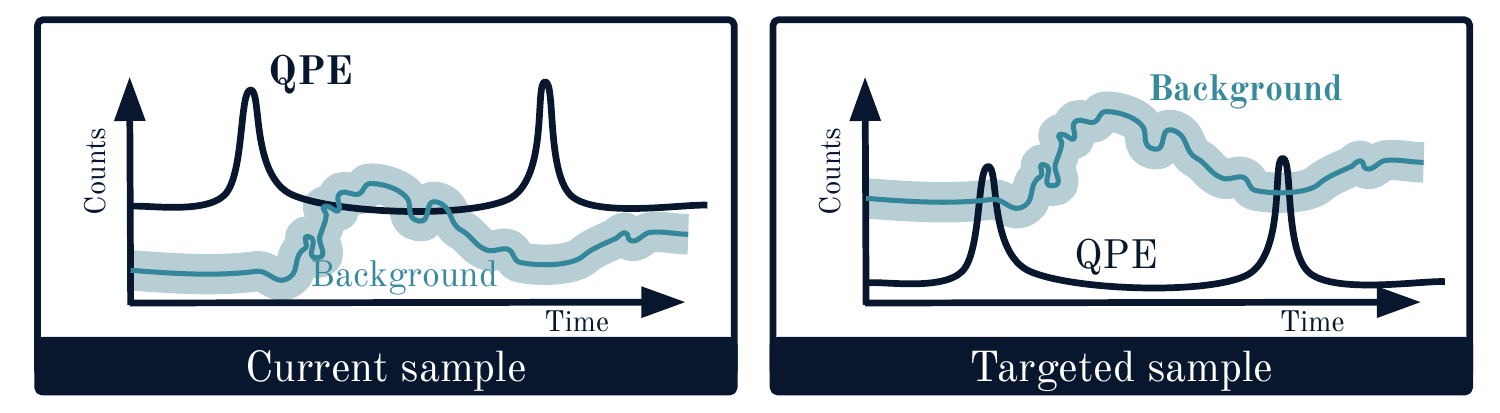}
    \caption{Illustration of the potential of EXOD for QPEs. The left panel depicts the current sample found in the archive, where the source is typically above the background at all times, or at least for a significant fraction of the eruptions. The right panel shows a much fainter sample of QPEs, where the source is only visible a few times above the background, such that it does not appear clearly in a stacked image.}
    \label{fig:EXOD2}
\end{figure*}

We mentioned in Sec. \ref{sec1} that archival searches of QPEs in the catalogs were unlikely to yield further candidates, as most of the easy-to-find candidates have already been reported. The method we suggest here relies on an alternative use of the archives, instead of the catalogs themselves, and makes use of the peculiar timing properties of QPEs. The general idea is relatively simple: for a given X-ray observation, catalogs detect sources by stacking the entirety of the exposure time \citep[and sometimes several observations, e.g.][]{traulsen_xmm-newton_2019}. This is ideal for constant sources, since both the background and sources counts will scale linearly with time, the signal-to-noise ratio will roughly scale as $\sqrt{t_{\rm exp}}$. However, in the case of transient sources (i.e. which are significantly above background levels only for a small amount of time compared to the total exposure), stacking the exposure will drown out a few frames of signal in an entire observation's worth of background (see Fig.~\ref{fig:EXOD1}). For a simple example, assuming a background rate of $10^{-3}$ photons s$^{-1}$, and a fast transient that appears at $1$ photons s$^{-1}$ for a 10s time window, over a total exposure time of 100ks (typical for XMM-Newton), the expected source counts are 10 photons, while the background counts are 100 for the total exposure, and 0.01 during the source's peak. The 10 photons of the source are of the same order of magnitude as the Poisson fluctuations of the background in the stacked observation, and as such will likely not be detected in the stacked exposure, while the source is clearly above the background in the correct time window. Performing time-resolved source detection would solve this issue, and allow to detect faint fast transients. This is the idea behind the EXOD pipeline \citep{pastor-marazuela_exod_2020, khan_exod_2025}.

EXOD takes an XMM-Newton event file as input, and distributes the photons in three-dimensional bins of 20"$\times$20"$\times t_{\rm bin}$, where $t_{\rm bin}$ is the user-specified time binning, which should correspond to the specific timescale of the type of transient the user is looking for. Several filtering steps are performed to remove most instrumental issues. EXOD then estimates the time- and space-variable background, and then compares the actual data bins to the estimated background. Any significant deviation from the expected background is either 1) a bright and variable source, 2) a transient source peaking above the background only for a few time bins, or 3) an instrumental artifact. By carefully filtering instrumental effects, this method thus allows us to retrieve variability in faint but detected sources, to detect bright transients that happened during instrumental Bad Time Intervals (which are usually discarded, but we can keep in our method), and more importantly to retrieve faint and fast transients that only appear for a few time bins \cite[for more details, see][]{khan_exod_2025}. 

We ran EXOD on the entirety of the XMM-Newton archive, corresponding to about 15 000 observations suitable to this method (i.e. in imaging mode), in three different time binnings (5s, 50s, 200s). It led to the discovery of 32 000 (respectively 4 000) transient events at a $3\sigma$ confidence level (resp. $5\sigma$). The soft band 0.2-2 keV accounts for half of these sources. This overwhelming number of potential new astrophysical objects of interest has barely been touched, and it might be an interesting new direction to search for QPEs.

Indeed, let us imagine a QPE source comparable to GSN~069, and presenting two eruptions over the course of $\sim$100ks of XMM-Newton observation  (see Fig.~\ref{fig:EXOD2}). GSN~069, which is at a distance of about 76 Mpc, peaked at a combined EPIC rate of about 5 photons s$^{-1}$ in its 2018 discovery observation. At a redshift 15 times larger (about 18 times more distant), this EPIC rate would become about $1.5\times10^{-2}$ photons s$^{-1}$, accounting for both the distance effect and the redshift pushing photons outside the XMM-Newton energy band. Considering the same eruption profile as GSN~069, this rate would translate into about 15 counts per peak, i.e. 30 counts in total. This is above the expected background in the corresponding time windows (about 10 counts in each 15mn time window), meaning EXOD would be able to detect it clearly. But it is negligible compared to the stacked background over the entire 100ks observation (about 1000 photons), which means that standard detection systems would have missed this source. Indeed, it is below the faintest sources in the standard catalog in terms of counts (typically $\sim$40 photons). Each individual peak would hence be detectable -- the repeated nature of the peaks, along with an extragalactic optical association, would be a smoking gun for a QPE identification. 

A dedicated QPE search using EXOD has not yet been performed. The general idea would be to look for transients with the following parameters
\begin{enumerate}
    \item soft emission, only in the 0.2-2 keV band;
    \item timescale ranging between $\sim$200s and $\sim$600s (over which typically this method is no longer advantageous compared to standard detection);
    \item showing at least two separate peaks in an observation;
    \item matching the position of a known galaxy.
\end{enumerate}
Preliminary studies have allowed to reveal an interesting candidate, \href{http://xmm-catalog.irap.omp.eu/source/208842501010030}{4XMM J235440.7-373019}, which was bright enough to be detected by the standard pipeline (with about 160 photons), and clearly shows two isolated peaks \citep[see][]{khan_exod_2025}. Its optical counterpart is unclear, and could be a star, in which case the transient would be a repeated stellar flare. Whatever its exact nature, it shows that QPE-like behaviour can be retrieved by EXOD. This would allow to perform blind search of faint archival QPEs in the XMM-Newton data, which would solve the various issues raised by the current methods (only archival data is used with no follow-up necessary, and such a blind search would be relatively free of confirmation bias, beyond the targeted timing and energy parameters). This comes at the cost of providing very faint QPE sources, for which standard spectral or timing studies are not possible, and for which follow-ups will be more difficult to motivate than for their brighter equivalents.

\section{Conclusions}\label{sec5}
In this work, we attempted a short review of the various methods to detect QPEs. The current sample has relied on either archival search in catalogs, blind search in eROSITA scans, or late-time X-ray monitoring of optical TDEs. These methods have several issues: the catalogs have been thoroughly exploited already; in the absence of eROSITA, no other mission has similarly ideal observing strategy for the serendipitous detection of QPEs; and following-up on TDEs requires a risky and costly X-ray monitoring, and introduces selection and confirmation biases in our sample. 

We suggest two avenues to mitigate these issues. First, following up on X-ray--detected TDEs (instead of optically-detected) would alleviate some of the risks of the TDE-based method, under the condition that proper X-ray transient detection systems are in place. It would still present the aforementioned selection biases. A second promising alternative method is based on a blind systematic search for faint QPEs, by using a time-resolved source detection method that has been developed for the XMM-Newton archive. This would be optimised to find very low count QPEs in the archive ($\sim$10-20 photons), peaking just above the background at the apex of the eruptions, and could detect sources similar to GSN 069 up to $\sim$20 times more distant in the Universe. Such a search would require a thorough investigation that has not yet been performed, but a proof-of-concept study has led to the discovery of a promising candidate.

\section*{Acknowledgments}
Softwares: \texttt{numpy} \citep{harris_array_2020}, \texttt{matplotlib} \citep{hunter_matplotlib_2007}, \texttt{cmasher} \citep{van_der_velden_cmasher_2020}.
EQ acknowledges support from the European Space Agency, through the Internal Research Fellowship programme. EQ acknowledges funding from the European
Union’s Horizon 2020 research and innovation programme under grant agreement number 101004168, the XMM2ATHENA project \citep{webb_xmm2athena_2023}. NW and RW acknowledge support of the CNES. GM acknowledges support from grants n. PID2020-115325GB-C31 and n. PID2023-147338NB-C21 funded by MICIU/AEI/10.13039/501100011033 and ERDF/EU.

\bibliography{references}%



\end{document}